\documentclass[amsmath, aps, prl, reprint, 
twocolumn, superscriptaddress]{revtex4-2}

\usepackage{amssymb}
\usepackage{lineno}
\usepackage{booktabs}
\usepackage{multirow}
\usepackage{soul, color, xcolor}
\usepackage{graphicx}
\usepackage{dcolumn}
\usepackage{bm}
\usepackage{hyperref}
\soulregister{\cite}7 
\soulregister{\ref}7 
\soulregister{\pageref}7 

\begin{document}


\title{Achieving Altermagnetism in Monolayer Holey Graphyne via Atomic Manipulation}


\author{Han-Bing Li}
\affiliation{Guangdong Basic Research Center of Excellence for Structure and Fundamental Interactions of Matter, Guangdong Provincial Key Laboratory of Quantum Engineering and Quantum Materials, School of Physics, South China Normal University, Guangzhou 510006, China}
\author{Zhi-Gang Shao}
\email{zgshao@scnu.edu.cn}
\affiliation{Guangdong Basic Research Center of Excellence for Structure and Fundamental Interactions of Matter, Guangdong Provincial Key Laboratory of Quantum Engineering and Quantum Materials, School of Physics, South China Normal University, Guangzhou 510006, China}
\affiliation{Guangdong-Hong Kong Joint Laboratory of Quantum Matter, Frontier Research Institute for Physics, South China Normal University, Guangzhou 510006, China}


\date{\today}

\begin{abstract}
Achieving altermagnetism (AM) in two-dimensional materials is crucial for advancing the development of novel spintronic devices. This study introduces an innovative strategy to realize AM in monolayer materials by adsorbing non-magnetic $sp$ impurity atoms to construct planar bridges. Using holey graphyne (HGY) as the research object, first-principles calculations reveal that B and S impurity atoms can modulate the local electronic structure and trigger a superexchange mechanism, inducing a collinear compensated antiferromagnetic ground state with pronounced altermagnetic properties. Among these, the B adsorption system exhibits the best magnetic performance, with a N$\rm \acute{e}$el temperature reaching approximately 210 K. This work offers a flexible and effective strategy to achieve AM in single-atomic-layer materials and $p$-electron systems.
\end{abstract}


\maketitle


\section{Introduction}

Antiferromagnets (AFMs) hold significant potential for spintronics due to their robustness against magnetic field disturbances, absence of stray fields, and ultrafast dynamics properties~\cite{1,2}. However, the spin degeneracy in conventional AFMs hinders the generation of spin-polarized currents~\cite{3}. Recently, a novel magnetic phase termed ``Altermagnetism (AM)" has attracted considerable attention. AM is characterized by a distinctive collinear spin arrangement, combining zero net magnetization with momentum-dependent spin splitting~\cite{4,5}. This electronic structure enables novel phenomena, including the crystal Hall effect, crystal thermal properties, spin-polarized currents, giant and tunneling magnetoresistance, providing fresh perspectives for spintronic device design~\cite{6,7,8,9,10,11,12}. The existence of AM has been experimentally verified by angle-resolved photoemission spectroscopy, which directly measures the non-relativistic spin-split electronic structure~\cite{13,14}.

To date, significant progress has been made in studying AM in three-dimensional systems, including MnTe and RuO$_2$~\cite{15,16,17}. In contrast, achieving AM in two-dimensional (2D) materials entails stricter symmetry constraints, resulting in a limited number of viable candidates~\cite{18}. Fortunately, recent studies have demonstrated that 2D materials, due to their low dimensionality and tunable properties, offer an ideal platform for exploring AM. AM in 2D materials can be realized through approaches like rotating van der Waals bilayers, bilayer stacking, or reverse stacking~\cite{19,20,21}. For example, the reverse-stacked structure of PtBr$_3$ and the twisted bilayer of NiCl$_2$ have been studied~\cite{20,21}. For monolayer structures, the collinear AFMs CrO has been predicted to exhibit significant momentum-dependent spin splitting~\cite{22}. Additionally, AM can also be realized by applying an electric field or using Janus structures, such as MnP(S,Se)$_3$ and RuF$_4$~\cite{23,24}. These approaches generate spin-split states via symmetry engineering, including introducing rotation axes or breaking spatial inversion symmetry. However, achieving AM in single-atomic-layer systems remains challenging, especially for $p$-electron systems, requiring strict symmetry control and precise regulation of local electronic structures.

Recently, the use of bridging B atoms for magnetic regulation in g-C$_3$N$_4$ has offered innovative solutions~\cite{25}. Building on these findings, this paper proposes a novel approach to realize AM in monolayer materials via atomic manipulation, selecting the 2D carbon allotrope holey graphyne (HGY) as the research object~\cite{26}. The unique nonlinear $sp$ bonding of carbon chains and the $\pi$-conjugated framework of HGY allow it to smoothly transition from $sp$ to $sp^2$ hybridization when adsorbing non-magnetic $sp$ impurity atoms (e.g., B, N, P, S), thereby minimizing structural deformation and maintaining system stability. Symmetry analysis and electronic structure calculations reveal that the unique topological structure of HGY enables the formation of collinear compensated spin arrangements at specific adsorption sites, achieving an altermagnetic state. This magnetic state is established by breaking time-reversal symmetry (TRS) while maintaining mirror symmetry. Furthermore, the study further reveals the origin of AM and the superexchange mechanism, and confirms the feasibility of achieving AM through magnetic exchange parameters and N$\rm \acute{e}$el temperature. This research, for the first time, achieves the design of AM at the single-atomic-layer scale, providing an efficient and universal scheme for magnetic engineering in the 2D limit.

\section{Model and Method}

The Vienna $ab$ $initio$ Simulation Package (VASP) was utilized for first-principles calculations~\cite{27,28,29}. All calculations were processed using the Projector Augmented Wave (PAW) method~\cite{30}. Wave functions were expanded using a plane wave basis set with a cutoff energy of 500 eV. The Perdew-Burke-Ernzerhof (PBE) functional within the Generalized Gradient Approximation (GGA) was used to describe the electron-ion exchange-correlation interactions~\cite{31}. A vacuum layer of 15 \AA{} perpendicular to the slab was set to avoid periodic interactions between adjacent layers. Van der Waals interactions were accounted for by applying the DFT-D3 correction~\cite{32}. All structures were fully relaxed until the total energy change was below $1\times10^{-5}$ eV and the Hellmann-Feynman force on each atom was below 0.01 eV/\AA{}, ensuring convergence. Structural optimization utilized a $6\times6\times1$ Monkhorst–Pack $k$-point mesh for Brillouin zone sampling~\cite{33}. An $11\times11\times1$ $k$-point mesh was applied to calculate the electronic structure and spin magnetic moments for improved accuracy. Ab initio molecular dynamics (AIMD) simulations employed a $2\times2\times1$ supercell structure and the NVT ensemble at 300 K, with a time step of 2 fs and a total simulation time of 5 ps. The adsorption energy $E_{\rm {ads}}$ was calculated using the following equation:
\begin{equation}
	E_{\rm {ads}} = E_{\rm {total}} - E_{\rm {sub}} - E_{\rm {atom}},
\end{equation}%
where $E_{\rm {total}}$, $E_{\rm {sub}}$, and $E_{\rm {atom}}$ represent the total energies of the atom adsorbed on HGY, HGY, and the impurity atom, respectively.

\section{Results and discussion}

\begin{figure}[!htb]
	\begin{center}
		\includegraphics[width=0.9\linewidth]{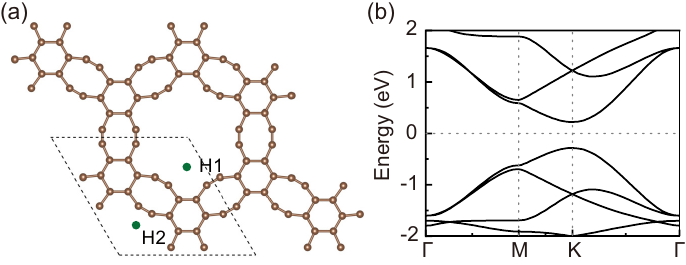}
	\end{center}
	\caption{(a) Geometric structure and adsorption sites of HGY. (b) Band structure of HGY. The dashed box highlights the unit cell of HGY.}
	\label{Fig1}
\end{figure}

Before studying the adsorption of impurity atoms on the HGY substrate, it is essential to analyze the structure and electronic properties of HGY. HGY is a 2D porous graphyne consisting of alternating six- and eight-membered carbon rings, as shown in Figure~\ref{Fig1}(a). The optimized lattice constants are $a = b = 10.85$ Å, aligning with prior reports~\cite{26}. HGY exhibits semiconducting properties with a direct band gap of 0.50 eV at the K point, as shown in Figure~\ref{Fig1}(b). HGY contains large carbon rings with six acetylene bonds, which can accommodate impurity atoms. Among these, H1 and H2 are located near the acetylene bonds and serve as ideal adsorption sites. After determining the adsorption sites, non-magnetic impurity atoms (B, N, P, S) are adsorbed at different sites in the HGY unit cell. It is found that the adsorption of impurity atoms at the H1 site bridges two eight-membered carbon rings, inducing AM in the system. Therefore, the subsequent discussion focuses on the electronic properties, magnetic characteristics, origins of magnetism, and phase transition temperatures of impurity atoms adsorbed at the H1 site.

\begin{figure}[!htb]
	\centering 
	\includegraphics[width=1.0\linewidth]{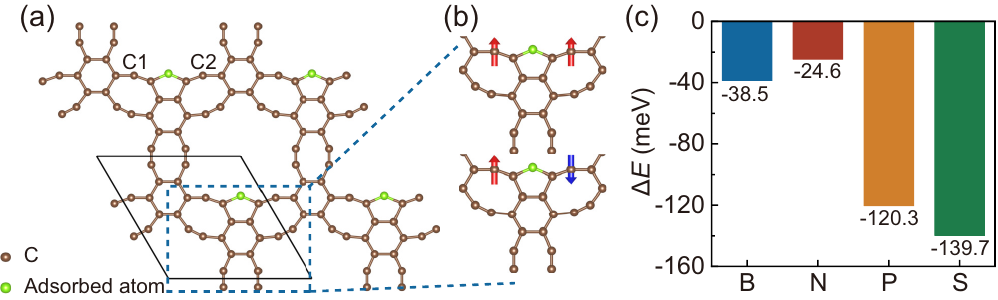}
	\caption{(a) Geometric structure of impurity atoms adsorbed on HGY. (b) FM (top) and AFM (bottom) configurations of the adsorption system. Arrows denote spin-up (red) and spin-down (blue), respectively. (c) Energy difference between the AFM and FM arrangements of the adsorption system.}
	\label{Fig2}
\end{figure}

\begin{figure*}[htb]
	\centering 
	\includegraphics[width=0.8\linewidth]{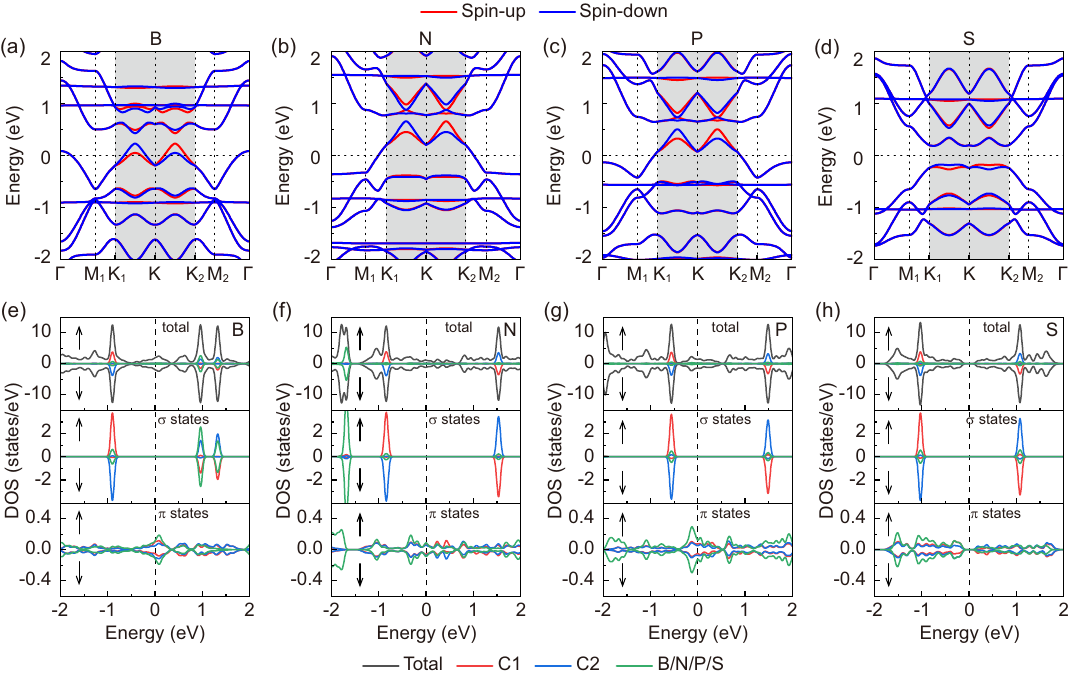}
	\caption{(a)-(d) Spin-polarized band structures and (e)-(h) DOS of $sp$ impurity atoms (B, N, P, S) adsorbed on HGY (without considering spin-orbit coupling). The direction of the arrows corresponds to the DOS for spin-up and spin-down.}
	\label{Fig3}
\end{figure*}

Initially, the stability of impurity atoms adsorbed on HGY under different magnetic states was evaluated to identify their ground state. As shown in Figure~\ref{Fig2}(a), the impurity atoms maintain a planar structure with HGY. The adsorption of impurity atoms generates unpaired electrons in the nearby two C atoms, thereby inducing magnetism. Consequently, ferromagnetic (FM) and antiferromagnetic (AFM) arrangements are considered, as shown in Figure~\ref{Fig2}(b). Figure~\ref{Fig2}(c) displays the energy difference between the two magnetic states, $\bigtriangleup E=E_{\rm AFM}-E_{\rm FM}$. The results indicate that all energy differences are negative, favoring AFM arrangement. Structural stability is vital for magnetic materials to achieve long-range spin order. The adsorption energies of impurity atoms on HGY were calculated to evaluate structural stability. The calculated adsorption energies of B, N, P, and S atoms on HGY are -3.47 eV, -4.08 eV, -2.59 eV, and -3.01 eV, respectively. Furthermore, the stability of high-temperature magnetic order in magnetic materials largely depends on lattice rigidity to resist thermal disturbances~\cite{34}. AIMD simulations were thus conducted to evaluate the thermal stability of the adsorption system. As shown in Figure S1, the B, N, and S adsorption structures remain stable at 300 K, without structural destruction or decomposition observed. In contrast, the P adsorption system exhibits significant energy fluctuations, with chemical bond breakage between P and C atoms. This is consistent with the adsorption energy results, confirming the instability of the P adsorption system.

\begin{figure}[!htb]
	\centering 
	\includegraphics[width=1.0\linewidth]{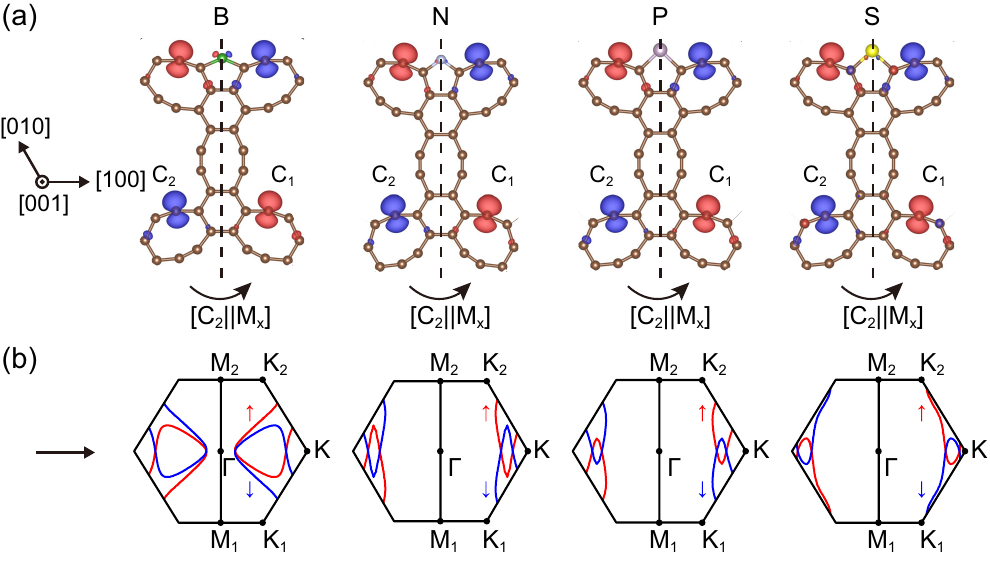}
	\caption{(a) Spin density diagram. The magnetic density of adjacent magnetic C sublattices exhibits twofold rotational symmetry around the $y$-axis and mirror symmetry about the symmetry axis perpendicular to the $xz$ plane. The dashed line marks the mirror symmetry axis. (b) Constant energy contour lines. The spin splitting properties of momentum points with mirror symmetry are opposite. The direction indicated by the arrows is the symmetry axis in momentum space.}
	\label{Fig4}
\end{figure}

Figures~\ref{Fig3}(a)-(d) present the non-relativistic spin-polarized band structures of the adsorption configurations in the AFM state. The results show that the B, N, and P adsorption systems transition to metals, while the S adsorption system remains its semiconductor properties due to the even number of electrons provided by the S atom. Notably, along the K$_1$-K-K$_2$ path, the bands for spin-up and spin-down split and are symmetric about the K point. The magnetic and non-magnetic C atoms form eight-membered rings with different orientations, connected by impurity atoms and carbon rings, resulting in two distinct magnetic sublattices that break spatial inversion symmetry. These two sublattices with opposite magnetic moments, are connected through real-space mirror reflection exchange, leading to the breaking of TRS in the band structure. The density of states (DOS) in Figures~\ref{Fig3}(e)-(h) show that the projected density of states (PDOS) for magnetic atoms C1 and C2 in the two spin channels is asymmetric, with significant spin polarization. However, the PDOS of C1 and C2 compensate each other, resulting in a symmetric total DOS and zero net magnetization. The spin density in Figure~\ref{Fig4}(a) indicates that C1 and C2 atoms are the main contributors to the spin charge density of the adsorbed system. The magnetic densities of the C1 and C2 sublattices can be mapped to each other through crystal mirror reflection. Additionally, in Figure~\ref{Fig4}(b), the constant energy contour lines near the Fermi level in momentum space show that the K-K$_1$ and K-K$_2$ paths are mirror-symmetric along the direction of arrow, with opposite spin splitting signs. Furthermore, $E(k,$\textuparrow$) \neq E(-k,$\textdownarrow$)$, further confirming the breaking of TRS. Based on the above rigorous spin symmetry analysis, it can be concluded that non-magnetic $sp$ impurity atoms adsorbed at the H1 site of HGY exhibit unique AM. Based on the above discussion, the methods for the emergence of AM in single-atomic-layer materials can be summarized as follows: (1) The crystal possesses mirror symmetry about the symmetry axis of the two carbon chains; (2) On the basis of minimal structural deformation due to adsorption, impurity atoms are adsorbed at the acetylene corner enclosed by the two carbon chains (located on the symmetry axis); (3) The adsorption of impurities leads to the appearance of magnetic atoms on both sides of the symmetry axis, and the entire system exhibits a spin antiparallel arrangement.

\begin{figure}[!htb]
	\centering 
	\includegraphics[width=1.0\linewidth]{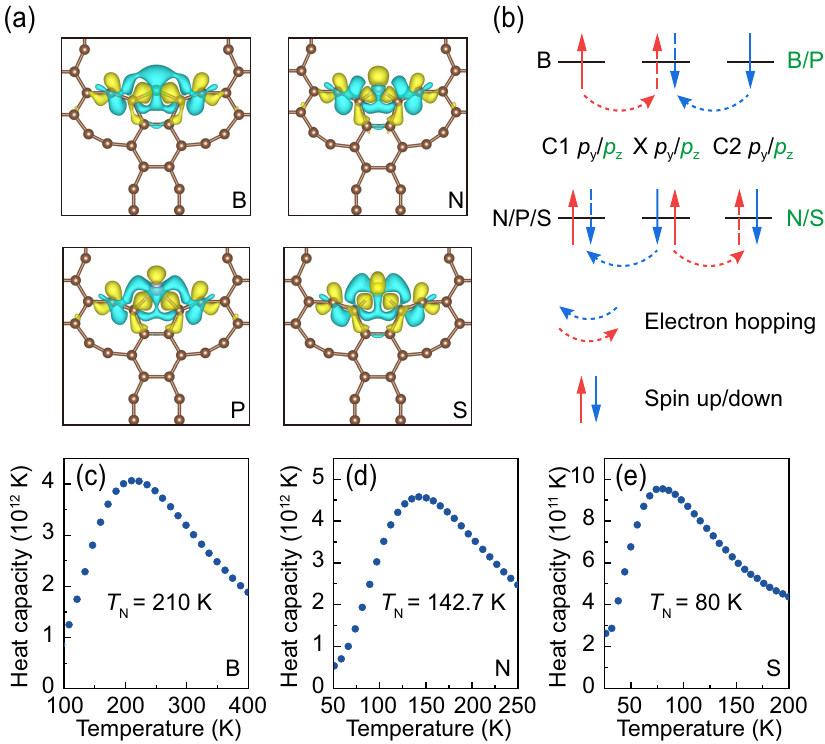}
	\caption{(a) Differential charge density and (b) superexchange mechanism of impurity atoms adsorbed on HGY. The black and green colors for B, N, P, and S correspond to the $p_y$ and $p_z$ orbitals, respectively. X represents the impurity atom. (c)-(e) Relationship between heat capacity and temperature of adsorption system.}
	\label{Fig5}
\end{figure}

To gain a deeper understanding of the magnetic origin of the adsorbed system, further calculations of the differential charge density were conducted for analysis. Figure~\ref{Fig5}(a) shows that impurity atoms bridge two eight-membered carbon rings in the plane, leading to a rearrangement of the electron cloud. There are distinct electron depletion regions near the two acetylene bonds adjacent to the impurity atoms, while electron accumulation occurs around the two C atoms that make up the acetylene bonds. This indicates that the impurity atom forms a $\sigma$ bond with one C atom, breaking the original $\pi$ bond and redistributing the local electron cloud. The $\sigma$ state energy level of the other C atom is disturbed. Its $p_y$ orbital participates in orbital hybridization to form a $\sigma$ state, generating unpaired electron states and local magnetic moments. Meanwhile, the $p_z$ orbital forms a new $\pi$ bond with a neighboring carbon atom. Notably, HGY has a special $sp$ hybridization bond angle, which facilitates the formation of $sp^2$ hybridization of C atoms and results in minimal structural deformation. By observing the differential charge density, it is found that there is a large charge depletion region above the B atom, in stark contrast to the N, P, and S atoms. This suggests that the B atom has an empty $sp^2$ hybridized orbital, while the N, P, and S atoms have a fully occupied hybridized orbital. There is charge depletion between the two C atoms of the six-membered ring in HGY, indicating that the large $\pi$ bond is disturbed. These two C atoms, along with the two $sp^2$ hybridized C atoms above and the impurity atom, form a five-membered ring, and their $p_z$ orbitals create new $\pi$ bonds. The five-membered ring shows significant charge accumulation. Its ring strain leads to a redistribution of the local electronic structure, potentially enhancing chemical reactivity and magnetism. Therefore, weak magnetic density is observed on the five-membered ring, as shown in Figure~\ref{Fig4}(a).

To further explore the relationship between electronic structure and magnetism, it is crucial to analyze the PDOS results from Figures~\ref{Fig3}(e)-(h) presented earlier. In the PDOS, the $\sigma$ state arises from the hybridization of $s$, $p_x$, and $p_y$ orbitals, while the $\pi$ state is composed of the delocalized $p_z$ orbitals. The results show that the DOS near the Fermi level is mainly contributed by $\pi$ electrons with weak spin polarization, indicating that $\pi$ electrons are not the main contributors to the spin magnetic moments of C1 and C2. The local magnetic moments of C atoms mainly originate from the spin polarization of electrons in the $\sigma$ state, confirming the presence of unpaired electrons in the $sp^2$ hybridized orbitals. The PDOS of impurity atoms shows that the $\sigma$ and $\pi$ states are symmetric in both spin channels, with no spin polarization observed. This indicates that the hybridized orbitals and $p_z$ orbitals of the impurity atoms do not contain unpaired electrons. Bader charge analysis shows that B, P, and S atoms lose 1.60, 1.03, and 0.20 electrons, respectively, while N atoms gain 0.99 electrons. The donors and acceptors of these electrons mainly come from the C atoms bonded to the impurity atoms. Combining this with the analysis of differential charge density, the $p_z$ orbitals of impurity atoms are either fully empty (B, P) or fully occupied (N, S), resulting in no net magnetic moment.

The PDOS indicates that near the Fermi level, impurity atoms exhibit similar energy levels and overlapping peaks with magnetic C atoms, suggesting the presence of coupling between atoms. Since C1 and C2 atoms are not directly bonded, the overlap implies that impurity atoms act as a medium, promoting indirect interactions between the electron spins of C1 and C2. AFM ordering arises from electron superexchange between magnetic and impurity atoms (C-X-C). According to the Pauli exclusion principle and the Goodenough-Kanamori-Anderson rules, superexchange allows electrons to virtually transition between C1 and C2 via impurities, establishing exchange coupling and lowering system energy~\cite{35,36}. Figure~\ref{Fig5}(b) illustrates that in the AFM configuration, the magnetic C atoms have half-filled orbitals, while impurity atoms have fully occupied or empty orbitals, enabling electrons to undergo virtual transitions between their $p_y$/$p_z$ orbitals. In the FM configuration, electron transitions would induce Pauli repulsive, raising the system energy. Therefore, magnetic C atoms are more inclined to AFM coupling. The DOS for the $\sigma$ state of the B system and the $\pi$ state of the B and P systems exhibit strong coupling near the Fermi level. This indicates that empty orbitals are more favorable for exchange coupling of electron transitions. Notably, exchange coupling is not limited by the arrangement of C-X-C. As shown in the differential charge density, regardless of the arrangement of C-X-C, the $p_y$ and $p_z$ orbitals of impurity and magnetic atoms are nearly parallel, with interatomic distances of $2.389\sim2.678$ Å, thus enabling orbital overlap. Therefore, non-magnetic atoms serve as bridges, inducing the generation of local magnetic moments in the system, facilitating their exchange, and thereby forming long-range magnetic order.

To quantitatively describe the exchange interactions and phase transition temperatures of the impurity-adsorbed HGY system, we analyzed the magnetic exchange coupling parameters using the 2D Heisenberg spin Hamiltonian. Detailed discussions are provided in the Supplemental Materials (S2). Notably, only the N and P systems show FM coupling for the next-nearest-neighbor exchange, deviating from the characteristics of AM. In contrast, the B adsorption system exhibits AM with high-temperature magnetic order (210 K). The corresponding heat capacity variations with temperature are shown in Figures~\ref{Fig5}(c)-(e). These findings highlight the critical role of superexchange mechanisms in determining the long-range magnetic interactions and N$\rm \acute{e}$el temperatures of the system.

\section{Conclusions}

This study proposes a design scheme for inducing AM in single-atomic-layer 2D materials via the adsorption of non-magnetic $sp$ impurity atoms, and systematically investigates the mechanism of achieving AM in HGY through the adsorption of impurity atoms B, N, P, and S. The research reveals the origin of the AM and its connection with the electronic structure and symmetry breaking of the system. The results indicate that after B and S atoms are adsorbed at specific sites on HGY, they can effectively induce a collinear compensated spin order with zero net magnetization and momentum-dependent spin splitting characteristics by bridging eight-membered carbon rings and triggering electron rearrangement. In contrast, N and P atoms fail to induce AM due to second-nearest-neighbor FM coupling. This method, for the first time, realizes the design of AM in single-atomic-layer materials, providing new ideas for the development of novel spintronic devices, and has two significant advantages. (1) Single-atomic-layer: Atomic bridging breaks the traditional limitation of requiring multilayer structures. This bridging method has been proven feasible in experiments~\cite{25}. (2) High tunability: By selecting different impurity atoms, the magnetic and electronic properties of the system can be flexibly regulated, providing extensive possibilities for designing multifunctional magnetic materials. Among them, the B, N, and P adsorption systems are metallic, while the S adsorption system is semiconducting. Notably, the B adsorption system demonstrates the best magnetic performance, achieving a N$\rm \acute{e}$el temperature of 210 K. These findings not only expand the research scope of AM but also provide a solid theoretical foundation and experimental guidance for achieving it in 2D $p$-electron systems.

\begin{acknowledgments}
This work was supported by the National Natural Science Foundation of China (Grant No. 52072132).
\end{acknowledgments}

\end{document}